\documentclass[12pt]{iopart}
\usepackage[dvips]{graphicx}
\usepackage{xfrac}
\usepackage{amssymb}
\newcommand{\be}{\begin{equation}}
\newcommand{\ee}{\end{equation}}
\newcommand{\bea}{\begin{align}}
\newcommand{\eea}{\end{align}}
\def\ket#1{|#1\rangle}
\def\bra#1{\langle#1|}
\def\V{V}

\def\Vs{\mathbb{C}^{2s+1}}
\def\Vi{V_i}

\def\refeq#1{(\ref{#1})}
\def\J{j}
\def\tj#1#2#3#4#5#6{\Big(\hbox{\begin{tabular}{ccc}$#1$&$#2$&$#3$\\$#4$&$#5$&$#6$\end{tabular}}\Big)}
\def\sj#1#2#3#4#5#6{\left\{\hbox{\begin{tabular}{ccc}$#1$&$#2$&$#3$\\$#4$&$#5$&$#6$\end{tabular}}\right\}}
\def\Jl{j_l}
\def\Jr{j_r}
\def\il{i_l}
\def\ir{i_r}
\def\til{i_l'}
\def\tir{i_r'}
\def\s#1#2{\sfrac{#1}{#2}}

\def\12{\frac{1}{2}}
\def\32{\frac{3}{2}}

\begin{document}

\title{Matrix product states for $su(2)$ invariant quantum spin chains}

\author{Rubina Zadourian$^1$, Andreas Fledderjohann$^2$, Andreas Kl\"umper$^2$}
\address{$^1$Max Planck Institute for the
Physics of Complex Systems, N\"othnitzer Stra{\ss}e 38,
01187 Dresden, Germany}
\address{$^2$Fachbereich C\ Physik, Bergische Universit\"at  Wuppertal,
42097 Wuppertal, Germany}


\begin{abstract}
A systematic and compact treatment of arbitrary $su(2)$ invariant spin-$s$
quantum chains with nearest-neighbour interactions is presented. The
ground-state is derived in terms of matrix product states (MPS). The
fundamental MPS calculations consist of taking products of basic tensors
of rank 3 and contractions thereof. The algebraic $su(2)$ calculations are
carried out completely by making use of Wigner calculus. As an example of
application, the spin-1 bilinear-biquadratic quantum chain is investigated.
Various physical quantities are calculated with high numerical accuracy of
up to 7 digits. We obtain explicit results for the ground-state energy,
entanglement entropy, singlet operator correlations and the string order
parameter. We find interesting crossover phenomena in the correlation lengths.
\end{abstract}


\section{Introduction}
Matrix product states (MPS) continue to attract scientific attention either as
a subject in its own right, or as realizations \cite{OstlundR95,RommerO97} in
density matrix renormalization group (DMRG) studies
\cite{White92}-\cite{Schollwoeck05}. In the condensed matter community, the
interest in such states developed with the discovery of AKLT models with exact
ground-states of the type of MPS with finite dimensional auxiliary space
\cite{AKLT87,AKLT88}, also called valence bond solids. Generalizations
including anisotropic models followed in \cite{Fannes89}-\cite{KolezhukM98b},
where especially in \cite{KlumperSZ91}-\cite{LangeKZ94} the formulation as
matrix product states was used. The models \cite{AKLT87,AKLT88}, despite their
fine-tuned interactions, were found to be representatives of `typical'
systems. The naturally appearing boundary degrees of freedom added even more
interest to the MPS states as they realize topological order
\cite{NijsRomm89,Tasaki91}.  On the practical side, the MPS allowed for a
variational treatment of the ground-state of a given Hamiltonian
\cite{OstlundR95,RommerO97,SchadschnZitt95}-\cite{KolezhukMY97} resulting in
the current understanding of DMRG techniques, see e.g.~\cite{Schollwoeck05}.
Finally, in the literature on integrable lattice systems 
MPS states appear where the local tensor
is called vertex operator \cite{JMMN92}-\cite{BoosGKS06}.

In this paper we are interested in the fundamental implementation of Lie
algebra symmetries in MPS with $su(2)$ as a most important example, however
with arbitrary spin-$s$ in the quantum space. By doing this, we are convinced
that any calculation can be performed more efficiently than without use of the
symmetry. To give an example of this reasoning we like to remind again of the
AKLT model \cite{AKLT87,AKLT88}. By use of a transfer matrix formalism, all
correlation functions of the state can be calculated from the eigenstates and
eigenvalues of the transfer matrix. In the case of AKLT, the transfer matrix
is nothing but a product of two $3j$ symbols. The eigenstates themselves are
nothing but $3j$ symbols, and the eigenvalues are $6j$ symbols. All of these
objects are `tabulated' and no actual calculation is needed, just the right
identification is necessary.

The use of Lie algebra symmetries in DMRG calculations has some history
\cite{Dukelsky98EPL}
-\cite{AWeichselbaum12}. In the paper \cite{FledderjohannKM11} two of the
current authors have started to formulate the local tensors of MPS and various
associated objects like transfer matrices in a $su(2)$ invariant manner by use
of Wigner calculus. The work \cite{FledderjohannKM11} was restricted to
spin-$1/2$ in the quantum space. In this paper we present the generalization
to arbitrary spin-$s$ and report on concrete applications to the spin-1
bilinear-biquadratic quantum chain which we investigate in a large part of the
Haldane phase. In addition to the generalization over \cite{FledderjohannKM11}
we managed to formulate less intricate graphical rules for the construction of
the basic components of MPS calculations. We believe that such transparent
rules are also essential for calculations in the higher dimensional case of
tensor network states, see for instance
\cite{KlumperNZ97}-\cite{EvenblyVidal10} and especially for variational
treatments \cite{Nishino00}-\cite{Xiang08}.

The paper is organized as follows. In Sect.~\ref{su2inv} we shortly summarize
the tensor calculus of MPS and introduce the $su(2)$ invariant local objects
based on Wigner's $3j$ symbols. Many objects are presented in diagrammatical
manner and are based on graphs with three-pointed vertices. The graphical
rules are formulated and explicit formulas for various kinds of transfer
matrices are given. We treat general nearest-neighbour couplings, entanglement
entropy and string order. In Sect.~\ref{Results} we present explicit results
from numerical variational calculations for the spin-1 bilinear-biquadratic
chain. The results are compared with numerical data of the literature.
In Sect.~\ref{Conclusion} we present our conclusions.

\section{Realization of $su(2)$ invariance}\label{su2inv}

We consider the class of matrix-product states
\be
\ket{\psi}=\Tr (g_1\cdot g_2\cdot ...\cdot g_L),\label{MPG}
\ee
where $g_i$ is a square matrix with some auxiliary (index) space $\V$. The
matrix entries of $g_i$ are elements of a local quantum space $\Vi$ which we
take as the $i$th copy of a $su(2)$ spin-$s$ space $\mathbb{C}^{2s+1}$.

The $su(2)$-invariance of $\ket{\psi}$ is guaranteed if the objects $g_i$ are
$su(2)$ invariant tensors in $V\otimes V^*\otimes\Vs $, where $V^*$ is the dual
space to $V$. We treat the case of a finite dimensional $V$, however with
arbitrary dimensionality. This space will consist of a direct sum over
irreducible spin-$\J$ representations where $\J=0, \12, 1, \32, 2,...$. Each
$\J$ may appear an arbitrary number of times $n_j$, in which case we label the
different orthogonal mupltiplets by an integer $i\in\{1,...,n_j\}$. The space
$V$ is spanned by orthogonal states $\ket{(\J,i),m}$ where the magnetic
quantum number $m$ varies from $-\J$ to $+\J$ in integer steps.

Let us consider in $V\otimes V^*\otimes\Vs$ any spin multiplet $(\J_1,i_1)$
from the first factor space, any spin multiplet $(\J_2,i_2)$ from the second
factor space, and the (only) spin multiplet $s$ of the third space.
Disregarding scalar factors, there is at most one way of coupling these
multiplets to a singlet state. The coupling coefficients are known as $3j$
symbols and the desired singlet is
\begin{eqnarray}
\nonumber
\hskip-2.5cm
g_{(\J_1,i_1),(\J_2,i_2)}^s:=\sum_{m_1,m_2,m}(-1)^{\J_2-m_2}
\tj{\J_1}{\J_2}{\J}{m_1}{-m_2}{m}\ket{(\J_1,i_1),m_1}\otimes
\bra{(\J_2,i_2),m_2}\otimes\ket{s,m}.\\
\label{g3j}
\end{eqnarray}
We are going to use graphical rules for the construction of $su(2)$ invariant
tensors like $g$. We associate with any graph consisting of three-pointed stars
(vertices) and closed or open directed lines (bonds) an $su(2)$-invariant
tensor:
\begin{itemize}
\item Bonds carry an angular momentum label $j$ and a label
  $m\in\{-j,...,+j\}$ where summation over $m$ is implied. The label $j$ is
  usually explicitly shown, $m$ is not. [In case we are dealing with several
  multiplets of type $j$ we number those with an integer $i$ and label
  the bond with $(j,i)$.]
\item Bonds evaluate to factors
\subitem $(-1)^{j-m}$ in case of a closed bond,
\subitem $\ket{j,m}$ in case of an open bond with outgoing arrow,
\subitem $(-1)^{j-m}\bra{j,m}$ in case of an open bond with ingoing arrow.
\item Vertices evaluate to a factor $\tj{j_1}{j_2}{j_3}{\pm m_1}{\pm m_2}{\pm
  m_3}$ where $(j_1,m_1)$, $(j_2,m_2)$, $(j_3,m_3)$ are read off from the
  bonds of the vertex in anti-clockwise manner. The sign of each entry $m_i$ 
  in the $3j$ symbol is determined by the arrow direction: $+$ and $-$ for
  `out' and `in'.
\end{itemize}
The number of open bonds is equal to the rank of the tensor. Note that due to
the symmetries of the $3j$ symbols an arrow on a closed bond with associated
label $j$ may be reverted resulting in a factor $(-1)^{2j}$.

\begin{figure}[h]
\begin{center}
\includegraphics[width=0.8\linewidth]{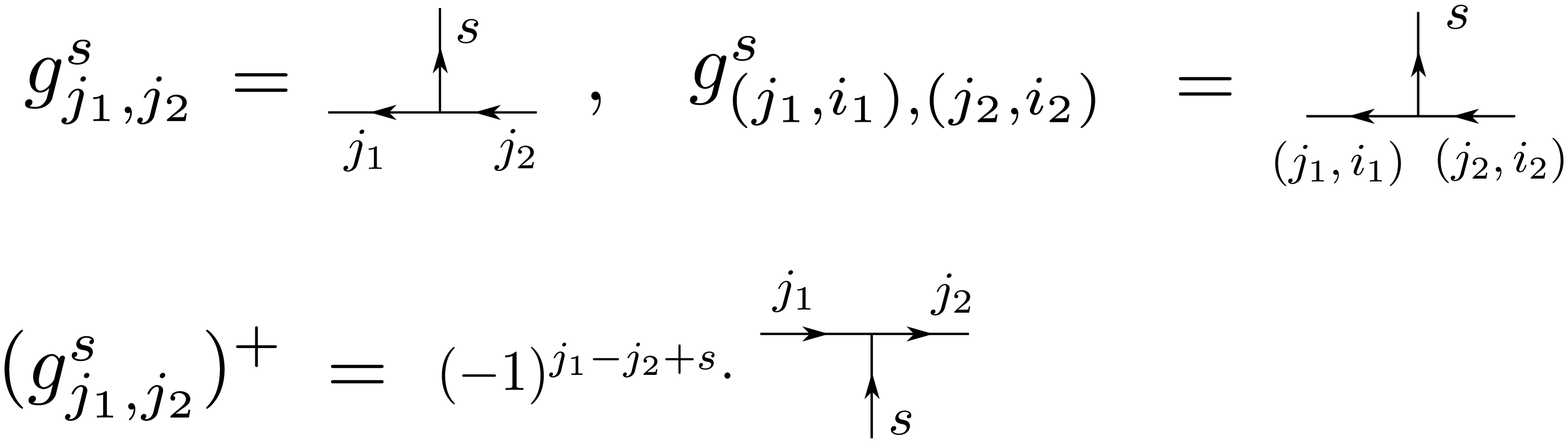}
\end{center}
\caption{Graphical depiction of the $su(2)$ invariant basic tensor
  $g_{\J_1,\J_2}^s$ and its dual
  $\left(g_{\J_1,\J_2}^s\right)^+$. Note that the sign factor
  $(-1)^{j_1-j_2+s}$ appearing in the graphical depiction of the dual can be
  dropped as it disappears in the products along the quantum chain.  In our
  applications there may be different copies of multiplets of the same type
  $j_1$ and $j_2$ associated with the bonds which will be explicitly noted
  like $g_{(\J_1,i_1),(\J_2,i_2)}^s$.}
\label{gandgbar}
\end{figure}

In Fig.~\ref{gandgbar} we show the graphical depiction of
$g_{(\J_1,i_1),(\J_2,i_2)}^{s}$ and its dual which is obtained by changing
ket-vectors into bra-vectors and vice versa.
A general $su(2)$ invariant tensor $g$ can be written as superposition of
these elementary singlets 
\be
g=\sum_{(\J_1,i_1),(\J_2,i_1)}A_{(\J_1,i_1),(\J_2,i_2)}\cdot g_{(\J_1,i_1),(\J_2,i_2)}^s,
\label{Gdurch3j}
\ee
with suitable coefficients $A_{(\J_1,i_1),(\J_2,i_2)}$. Note that $s$ does
not appear as index of $A$ as $s$ is always fixed and unique 
(for this reason no $i_3$ has been introduced either). Due to the symmetry of 
$3j$ symbols with respect to exchange of two columns
\be
\tj{\J_3}{\J_2}{\J_1}{m_3}{m_2}{m_1}=(-1)^{\J_1+\J_2+\J_3}
\tj{\J_1}{\J_2}{\J_3}{m_1}{m_2}{m_3},\label{transpose3j}
\ee
we conclude that
\be
A_{(\J_1,i_1),(\J_2,i_2)}=\pm (-1)^{\J_1+\J_2+s}A_{(\J_2,i_2),(\J_1,i_1)},
\label{transposeA}
\ee
with globally fixed $\pm$, is a sufficient condition for parity
invariance. Note that $\J_1+\J_2+s$ is always integer.

Also note that only few combinations ${\J_1}, {\J_2}, {s}$ need to be
considered: if the triangle condition $|\J_1-\J_2|\le s\le|\J_1+\J_2|$ is
violated, the three multiplets can not couple to a singlet.

\subsection{Norm and transfer matrix}

\begin{figure}[h]
\begin{center}
\includegraphics[width=0.6\linewidth]{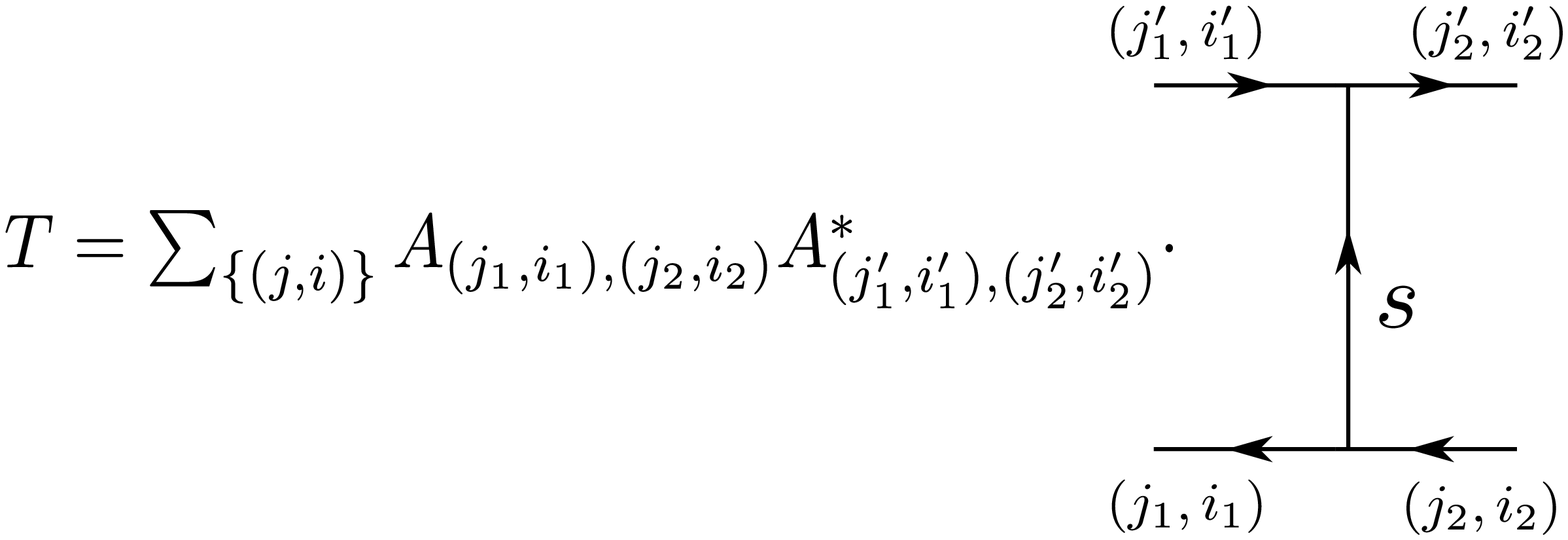}
\end{center}
\caption{Definition of the transfer matrix $T$. The graph
  appearing in the definition evaluates according to the given rules to a
  $su(2)$ invariant tensor. The sum extends over all four pairs of
  labels $(j_1,i_1), (j_2,i_2), (j_1',i_1'), (j_2',i_2')$.}
\label{TransferM}
\end{figure}

Next, we want to calculate the norm $\bra{\psi}\psi\rangle$ and the
expectation value of the Hamiltonian $\bra{\psi}H\ket{\psi}$ in the
thermodynamic limit. The computation leads to
\be
\langle\psi\ket{\psi}=\Tr (g_1^+g_1\cdot g_2^+g_2\cdot ...\cdot g_L^+g_L),\label{norm}
\ee
where $g^+$ $\in V^*\otimes V\otimes(\Vs)^*$ is the dual of $g$ $\in
V\otimes V^*\otimes\Vs$ and the contraction over the third space is
implicitly understood in $g^+ g$. Hence $T:=g^+ g$ is a linear map 
$V\otimes V^*\to V\otimes V^*$. A quite explicit expression for $T$ is given
in a mixed algebraic and graphical manner in Fig.~\ref{TransferM}.





For the computation of the norm we employ the
transfer matrix trick yielding for the r.h.s. of \refeq{norm}
\be
\langle\psi\ket{\psi}=\Tr T^L =\sum_{\Lambda}\Lambda^L,\label{puretransfermatrix}
\ee
where the sum is over all eigenvalues $\Lambda$ of $T$. Obviously, in the 
thermodynamic limit only the largest eigenvalue(s) $\Lambda_0$ contribute. The
corresponding left and right eigenstates are denoted by $\bra{0}$ and $\ket{0}$.

The computation of the leading eigenvalue is facilitated by the singlet nature
of the leading eigenstate. There are not many independent singlet states in 
$V\otimes V^*$. A $(\J,i)$ multiplet in $V$ and a $(\J',i')$
multiplet in $V^*$ couple to a singlet iff $\J=\J'$ (with arbitrary $i$
and $i'$). The (normalized) singlet is given by
\be
\sigma(\J;i,i'):=\frac1{\sqrt{2\J+1}}
\sum_{m=-\J}^\J\ket{\J,i,m}\otimes\bra{\J,i',m}.
\ee
Graphically, this singlet is depicted by a link carrying an arrow pointing
from the $V^*$ to the $V$ space, a label $j$ in the middle and $i$, $i'$ at
the respective ends.

\begin{figure}[h]
\begin{center}
\includegraphics[width=0.75\linewidth]{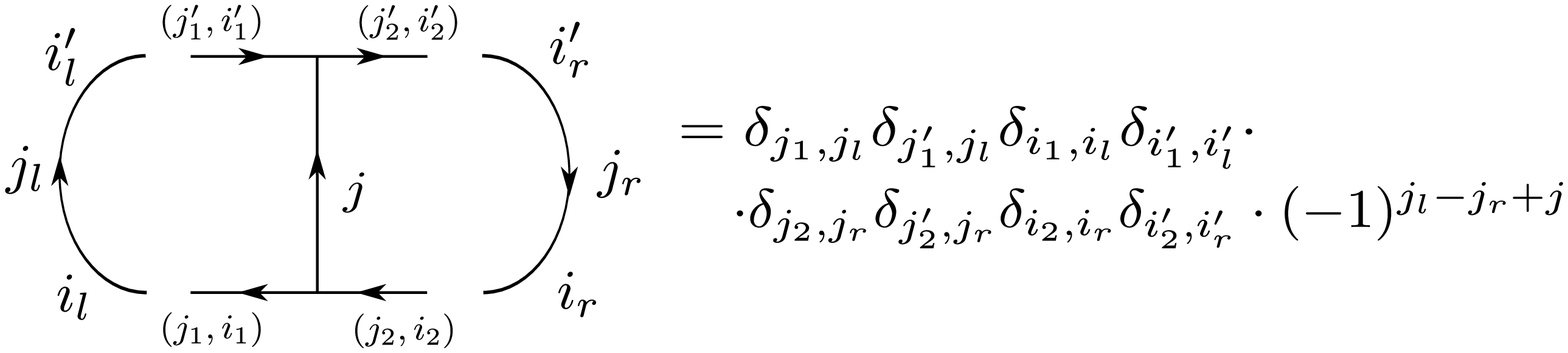}
\end{center}
\caption{Multiplication of fundamental tensor with unnormalized singlet states.}
\label{MatrixElem}
\end{figure}

The action of the transfer matrix $T$ onto singlets produces singlets with
matrix elements 
\be
\bra{\Jl;\il,\til}T\ket{\Jr;\ir,\tir}=
\frac{(-1)^{j_l-j_r+s}}{\sqrt{(2\Jl+1)(2\Jr+1)}}
A_{(\Jl,\il),(\Jr,\ir)}A^*_{(\Jl,\til),(\Jr,\tir)}.\label{Telements}
\ee
For this calculation the identity depicted in Fig.~\ref{MatrixElem} has been
used (where for the derivation of (\ref{Telements}) $j=s$ is to be used).
The total dimension of $V\otimes V^*$ is $\Big[\sum_\J
n_\J(2\J+1)\Big]^2$, but the singlet subspace is much lower dimensional:
$\sum_\J n_\J^2$.  Due to the still high dimensionality, the diagonalisation
in the singlet space has to be done numerically.

In order to give full meaning to the graphical operations in
Fig.~\ref{MatrixElem} we complement our rules by
\begin{itemize}
\item Isolated (directed) bonds carry an angular momentum label $j$
  and evaluate to a product of the factors 
\subitem $\ket{j,m}$ for the end with outgoing arrow,
\subitem $(-1)^{j-m}\bra{j,m}$ for the end with ingoing arrow.
\item Concatenations of two or more graphs correspond to contractions over
  pairs of open
  bonds and naturally lead to identifications of the labels that are carried
  by the joined bonds.
\end{itemize}
Note that these rules are consistent with the above rule for the factor
contributed by a closed bond. Also note that indices $i$ identifying different
multiplets of the same type $j$ appear on the bonds and usually pertain to the
entire bond. The only exception are the singlet states represented by arcs
with possibly different labels $i$ and $i'$ at the ends.

\subsection{Nearest-neighbour couplings}

\begin{figure}[t]
\begin{center}
\includegraphics[width=0.35\linewidth]{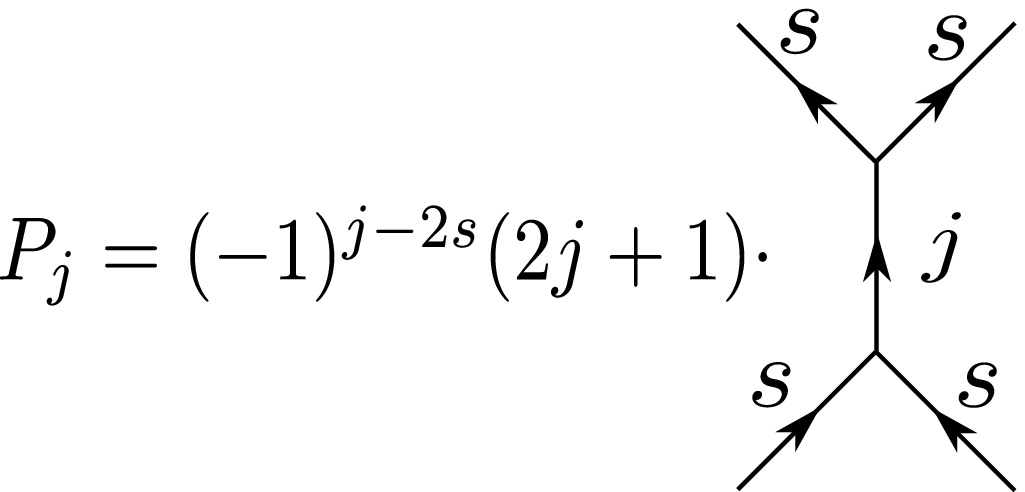}
\end{center}
\caption{Depiction of the projection operator on two spin-$s$ spaces onto the
  total spin-$j$ subspace.}
\label{projection}
\end{figure}

\begin{figure}[h]
\begin{center}
\includegraphics[width=0.75\linewidth]{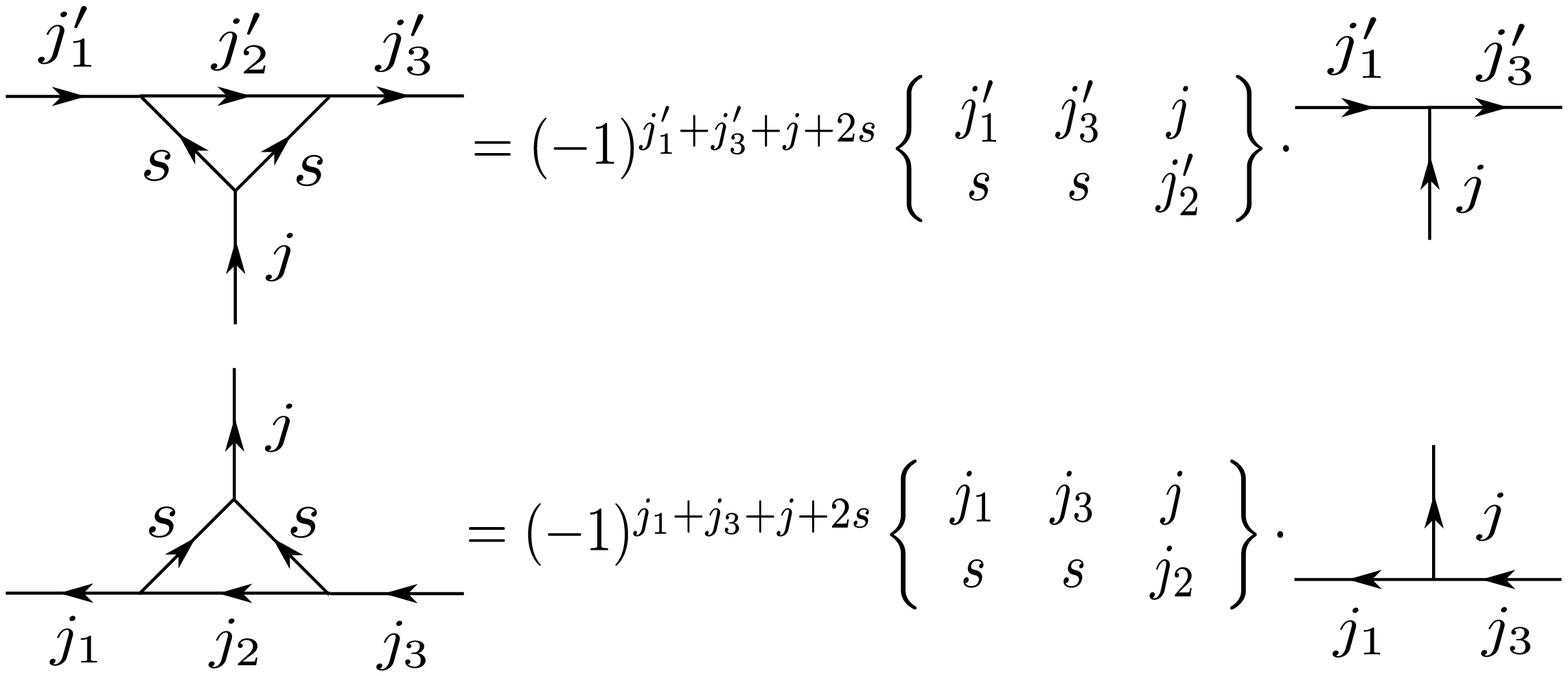}
\end{center}
\caption{Reduction of `triangles to stars' by use of fundamental properties of
  $3j$ and $6j$ symbols}
\label{Reduction}
\end{figure}

\begin{figure}[h]
\begin{center}
\includegraphics[width=0.95\linewidth]{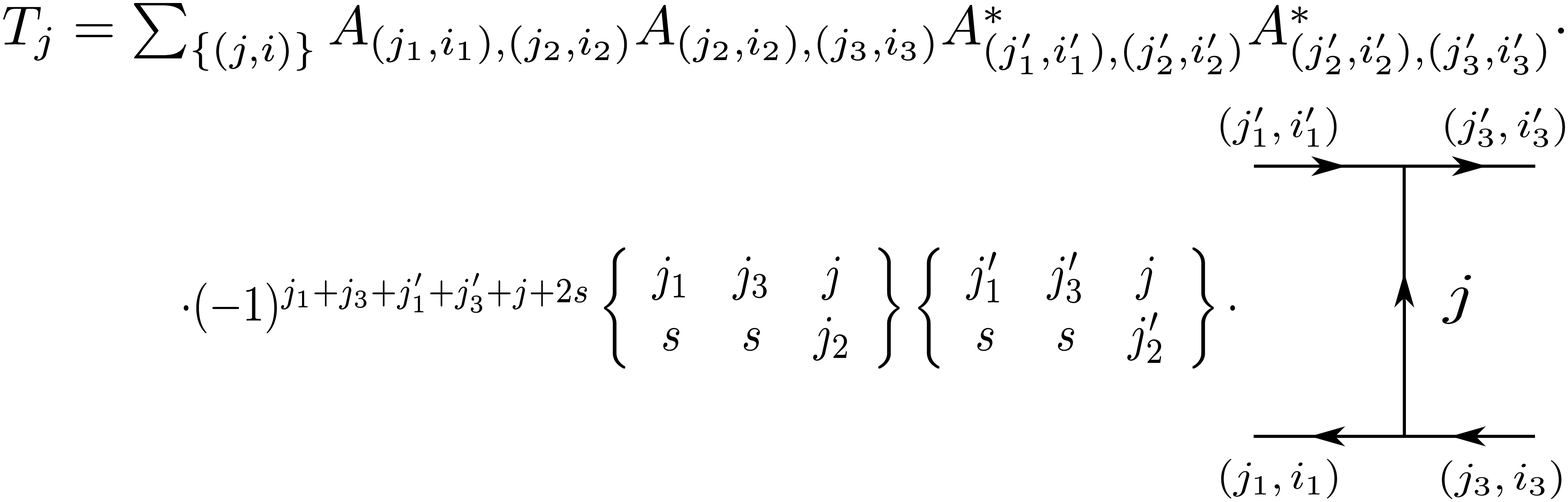}
\end{center}
\caption{The explicit expression for the modified transfer matrix $T_j$
  accounting for the action of the projection operator $P_j$ on two lattice
  sites. The summation extends over all six pairs of the labels $(j_1,i_1),
  (j_2,i_2), (j_3,i_3), (j_1',i_1'), (j_2',i_2'), (j_3',i_3')$.}
\label{TransferMod}
\end{figure}

We are interested in the spin-$s$ Heisenberg chain with most general
nearest-neighbour interaction. The corresponding Hamiltonian can be written in
two alternative ways
\be H=\sum_{l=1}^L Q(\vec S_l
\vec S_{l+1})=\sum_{l=1}^L \sum_{j=0}^{2s-1}a_j P_j \qquad(+\hbox{const.})
\ee
either with a polynomial $Q$ of degree $2s$ applied to the scalar product of
the nearest-neighbour spin vectors, or as a superposition of $2s$ many
projection operators $P_j$ onto nearest-neighbour spin multiplets $[j]$.
The operator $P_j$ is given in terms of $3j$ symbols like
\begin{eqnarray}
P_j=(2j+1)&\sum_{m_1,m_2,m,m_1',m_2'}&
\tj{s}{s}{j}{m_1'}{m_2'}{m}\tj{s}{s}{j}{m_1}{m_2}{m}\nonumber\\
&&\cdot\ket{s,m_1'}\otimes\ket{s,m_2'}\bra{s,m_1}\otimes\bra{s,m_2},
\end{eqnarray}
and is graphically depicted in Fig.~\ref{projection}.
We want to determine the
matrix-product state with minimal expectation value of the total Hamiltonian
$H=\sum_l h_l$. Due to translational invariance the
expectation value of a single projector $P_j$ acting on sites 1 and 2 needs to
be calculated leading to
\be 
\bra{\psi} P_j\ket{\psi}=\Tr (T_j\underbrace{T\cdot ...\cdot T}_{L-2\ \mbox{times}})
=\Lambda_0^{L-2}\bra{0}T_j\ket{0}.
\label{modiftransfermatrix} 
\ee 
$T_j$ is a modified transfer matrix acting in $V\otimes V^*$, $\ket{0}$ is the
(normalized) leading eigenstate of the transfer matrix $T$ and we kept the
only term that dominates in the thermodynamical limit. Hence
\be
\frac{\bra{\psi} P_j\ket{\psi}}{\langle\psi\ket{\psi}}
=\Lambda_0^{-2}\bra{0}T_j\ket{0}.
\ee
For the explicit expression of $T_j$ we use the calculus of Wigner
symbols. $T_j$ is obtained from the multiplication of $P_j$ with two $g$
tensors and two $g^+$. The underlying graph contains two triangles that allow
for an explicit evaluation of the internal sums over the `magnetic' quantum
numbers. In this way, triangles are reduced to vertices times $6j$
symbols. Details of these calculations are shown in Fig.~\ref{Reduction}.
The final result of these calculations is given in Fig.~\ref{TransferMod}.

\subsection{Entanglement }\label{Entang}
Next, we consider a finite segment of length $l$ inside a very long quantum
chain. The reduced density matrix $\rho_l$ is obtained from the total density
matrix $\rho$ by taking the trace over all local spaces except for 
those at sites $1,..., l$
\be
\rho_l:=\Tr_{\tiny\mbox{partial}}\rho.
\ee
This object can be written in terms of basis states on the $l$ sites
\begin{eqnarray}
\ket{\alpha,\beta;l}&:=(g_1 g_2\cdot ...\cdot g_l)_{\alpha,\beta}\,,\nonumber\\
\bra{a,b;l}&:=(g^+_1 g^+_2\cdot ...\cdot g^+_l)_{a,b}\,.
\end{eqnarray}
Note that these states are usually not orthonormalized, but satisfy
\be
\bra{a,b;l}{\alpha,\beta;l}\rangle\simeq_{l\to\infty}\Lambda_0^l
\underbrace{\langle {\alpha,a}|0\rangle}_{=:R_\alpha^a} \underbrace{\langle 0|{\beta,b}\rangle}_{=:L_\beta^b},
\ee
where $\bra{0}$ and $\ket{0}$ are the left and right leading eigenstates of the
transfer matrix $T$, and we have defined the matrices $R$ and $L$.

The matrix elements of the reduced density matrix can then be given in the
form
\be
\bra{a,b;l}\rho_l\ket{\alpha,\beta;l}=(RL^TR)^a_\alpha\cdot(LR^TL)^b_\beta.
\ee
Hence, the reduced density matrix in an orthonormalized basis is isomorphic to
$RL^T\otimes LR^T$. Furthermore, the entanglement entropy is twice the
entropy of the matrix $RL^T$.

\subsection{Hidden order}\label{Hidden}
Next, we are interested in the computation of long-range string
order \cite{NijsRomm89,Tasaki91}, i.e.  `hidden order'. Due to symmetry, the
standard two-point spin-spin-correlators decay to zero for large separations
$l$ of the operators $S_0^z$ and $S_l^z$.  However, the expectation value of
the spin operators times ${\cal{O}}_{1,l-1}:=\exp(\pi\mbox{i}\sum_{1\le j\le
  l-1}S_j^z)$ does not necessarily vanish. It is known to be non-zero for the
Haldane phase of the Heisenberg chain. By use of transfer matrices we find
\be
\langle S_0^z \,{\cal{O}}_{1,l-1}\, S_l^z\rangle=
\Lambda_0^{-l-1}\bra{0}T^z (T')^{l-1} T^z\ket{0},
\ee
where the modified transfer matrices $T^z$ and $T'$
are defined similar to the standard
tranfer matrix $T$ with the additional action of the spin operators on the
physical space: $T^z:=g^+ S^z g$ and $T':=g^+\exp(\pi\mbox{i}S^z) g$.
In case of $T'$ we may apply an important simplification rewriting the action
of $\exp(\pi\mbox{i}S^z)$ on the third component of $g$ as the action
of $\exp(\pi\mbox{i}(-S^z_1+S^z_2))$ on the first two components of $g$. This
is due to $su(2)$ (actually $u(1)$) invariance: on the r.h.s.~of (\ref{g3j})
only terms with $m_1-m_2+m=0$ contribute. Now these operators on the horizontal
bonds cancel pairwise in the product of the $l-1$ many $T'$ transfer matrices,
except for one factor on the left and one on the right end. These factors may
then be associated in the general bookkeeping with the $T^z$ resulting in
\be
\langle S_0^z \,{\cal{O}}_{1,l-1}\, S_l^z\rangle=
\Lambda_0^{-l-1}\bra{0}T^z_r T^{l-1} T^z_l\ket{0},
\ee
where $T^z_{r,l}$ are identical to $T^z$ with the additional action of 
$\exp(-\pi\mbox{i} S^z_2)$ and $\exp(\pi\mbox{i}S^z_1)$ in the $r$ and $l$
case. For large separation we find the leading contribution
\be
{\cal{O}}_2:=\lim_{l\to\infty}\langle S_0^z \,{\cal{O}}_{1,l-1}\, S_l^z\rangle=
\Lambda_0^{-2}\bra{0}T^z_r\ket{0}\bra{0}T^z_l\ket{0},
\ee
just from knowing the leading eigenstate of $T$.

Working with the modified transfer matrices leads to the insertion of a vertex
($3j$ symbol) into the vertical line in the graphical depiction in
Fig.~\ref{TransferM} as well as associating minus signs with the lower left
and right bonds. The additional vertex has two bonds with spin-$s$ and a third
label with spin-$1$ (and $m=0$ for dealing with the $z$ component of the spin
operator). 

We need matrix elements of $T^z_{r,l}$ in the singlet subspace. This result
looks like (\ref{Telements}) with two additional factors on the r.h.s. A
factor $\sqrt{s(s+1)(2s+1)}$ stems from treating the spin operators by $3j$
symbols and takes care of the normalization $(\vec S)^2=s(s+1)$.  The second
factor originates from the summation over internal $m$ variables of a graph
where a triangle is reduced to a star yielding a $6j$ symbol. For $T^z_{r}$ we
obtain the factor
\be
 (-1)^{j_r}\sj{j_r}{j_r}{1}{1}{1}{j_l}\sum_m\tj{j_r}{j_r}{1}{m}{-m}{0},
\ee
where the last sum evaluates to $\sqrt{(2j_r+1)/(4j_r(j_r+1))}$ for half odd
integer $j_r$ and simply $0$ for integer $j_r$.

\section{Results}\label{Results}

For the nearest-neighbour spin-$1$ quantum chain with bilinear and biquadratic
interaction
\be 
H=\sum_{l=1}^L \left(\vec S_l\vec S_{l+1})+\alpha \big(\vec S_l\vec S_{l+1}\big)^2\right)
\ee
we performed calculations for $-1\le\alpha\le 1/3$. In the auxiliary space we
used alternatively purely integer spin multiplets and purely half-odd integer
spin multiplets. In the diagrams we show results for $n_{\s12}=5$,
$n_{\s32}=4$, $n_{\s52}=2$, $n_{\s72}=1$ ($n_s$=0 for $s>7/2$) where $n_s$ is
the number of independent spin-$s$ multiplets.  This corresponds to a
46-dimensional auxiliary space.  Taking into account a gauge freedom already
described in \cite{FledderjohannKM11} we have to deal with 41 variational
parameters. The actual calculations have been carried out in Maple 13. The
minimum in energy was found by the built-in optimization routine and also --
with same result -- by a simple gradient procedure.

For $\alpha=0$ we found a ground-state energy per site of $e_0=-1.4014838$
which is off by (only) $2\cdot 10^{-7}$ from the best known numerical value of
$-1.401484038971$ \cite{WhiteHuse93}. Our result for the string order
parameter is ${\cal{O}}_2=0.37433$ to be compared with $0.374 325 09$ by
\cite{WhiteHuse93}. We obtain for the entanglement entropy $S=2.778$ and for the
correlation length of singlet operators $\xi=1.943$. Our value for the
entropy compares well with that reported in \cite{DengKQ14}.

At $\alpha=-1$ the model is critical and the accuracy of the data in the above
approximation is worse. Still the results for the ground-state energy
$e_0=-3.992$, the expectation values of the projectors onto
nearest-neighbour singlets and triplets $\langle P_0\rangle=0.572$ and
$\langle P_1\rangle=0.094$ deviate only by $10^{-3}$ from the exactly known
values $e_0=-4$, $\langle P_0\rangle=0.5732747261...$ and
$\langle P_1\rangle=0.093391941...$ \cite{KNSuzuki13}.

At $\alpha=1/3$ our results reproduce the known analytical facts about the
AKLT model \cite{AKLT87,AKLT88}. Here only the correlation parameter of a
single $s=1/2$ multiplet in the auxiliary space with itself is enough to
realize the exact ground state with $e_0=-2/3$, string order parameter
${\cal{O}}_2=4/9$ and entanglement entropy $S=\log_2 4=2$. All other
correlation (variational) parameters are 0. The state with a single spin-1/2
multiplet on each bond is of course the VBS state where the $3j$ symbol at the
vertex produces a spin-1 by symmetrization, and the product of two vertices in
the auxiliary space (contraction) corresponds to antisymmetrization of two
spin-1/2 objects.

In the course of our numerical experiments we found that the same state can be
constructed also by integer spin multiplets in the auxiliary space.  This can
be understood analytically: One needs for instance one spin-0 and one spin-1
multiplet. The spin-0 and spin-1 couple with weight $1/\sqrt{6}$ and the
spin-1 with itself with `weight' $-1/3$.

\begin{figure}[h]
\begin{center}
\includegraphics[width=0.7\linewidth]{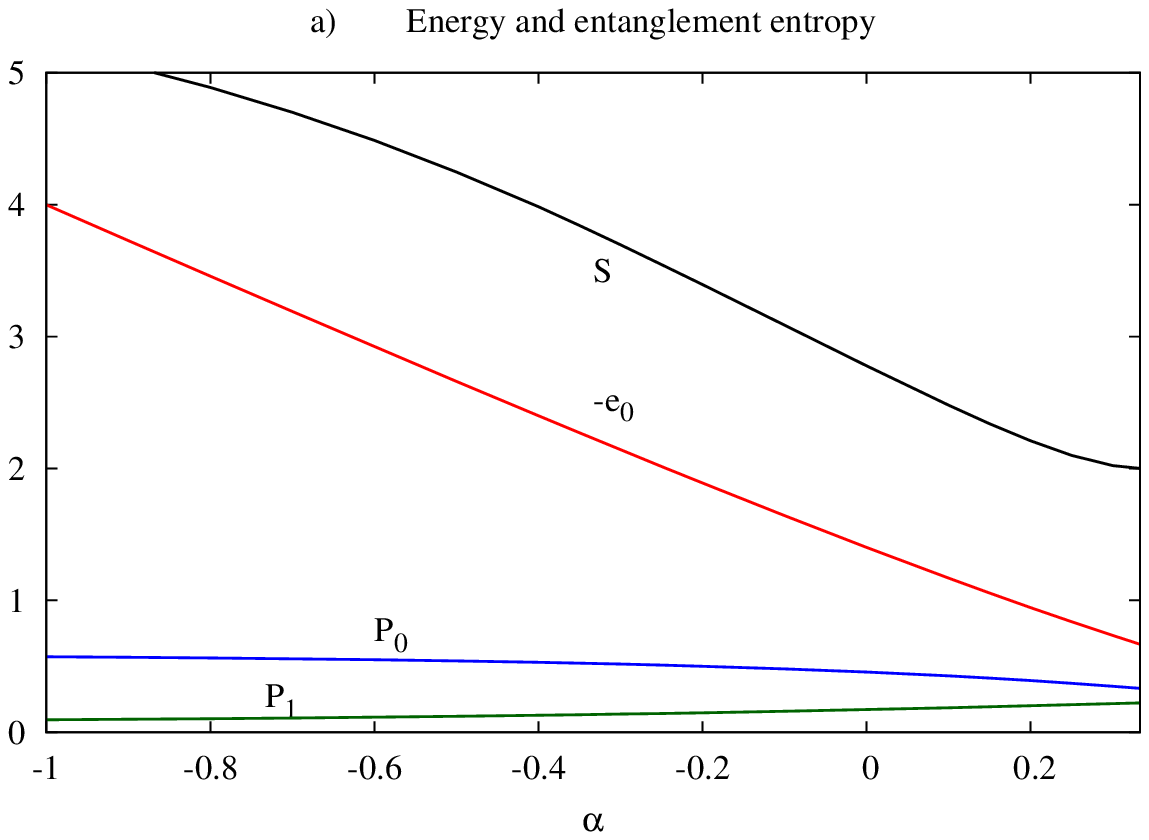}\vskip0.2cm
\includegraphics[width=0.7\linewidth]{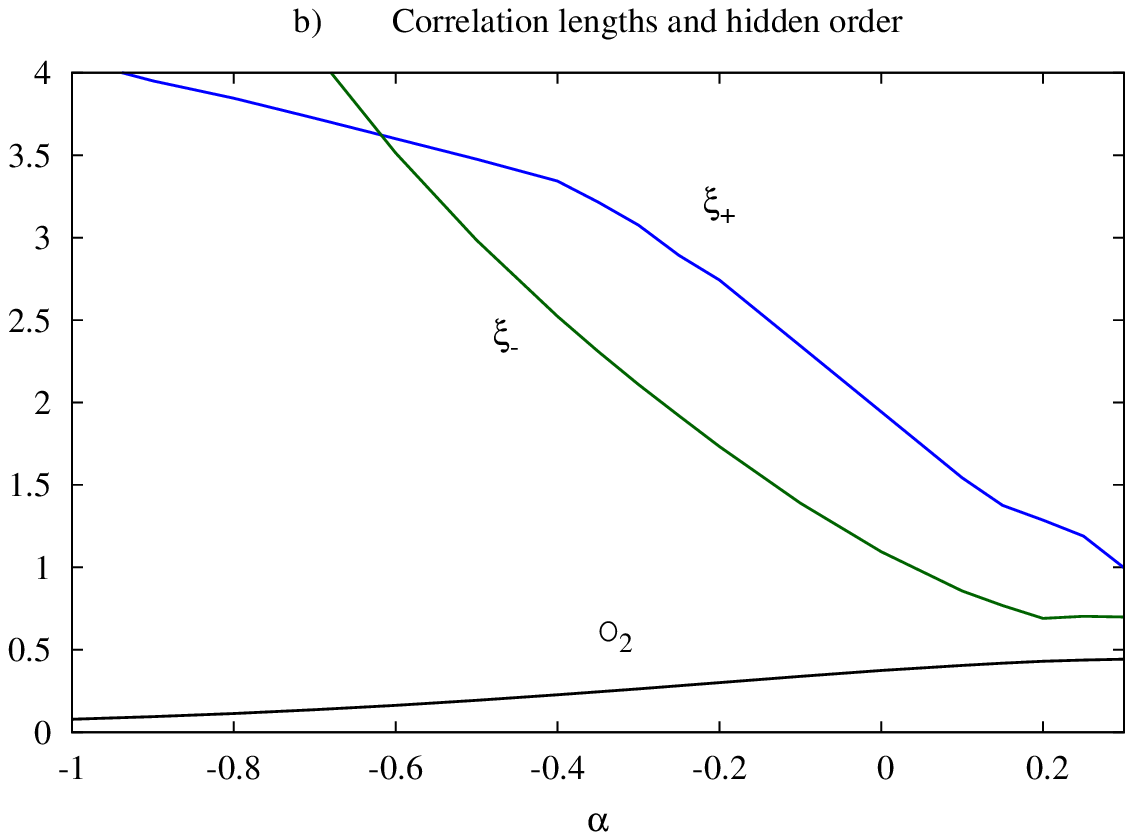}
\end{center}
\caption{a) Plot of (minus) the ground-state energy $e_0$ and of the
  entanglement entropy $S$ versus the interaction parameter $\alpha$. Also
  shown are the expectation values of the projection operators $P_0$ and $P_1$
  onto nearest-neighbour singlets and triplets.  b) Plot of the correlation
  lengths $\xi_\pm$ for singlet operators. Here $\xi_+$ describes the
  uniformly decaying exponential contribution, and $\xi_-$ describes the
  oscillating contribution. Note the crossover of the correlation lengths at
  $\alpha_c\simeq-0.6$. Also shown is the string order parameter ${\cal{O}}_2$
  where the small finite result at $\alpha=-1$ is due to the finiteness of the
  number of multiplets in our MPS calculations.}
\label{graphdimeral}
\end{figure}

\begin{figure}[h]
\begin{center}
\includegraphics[width=0.7\linewidth]{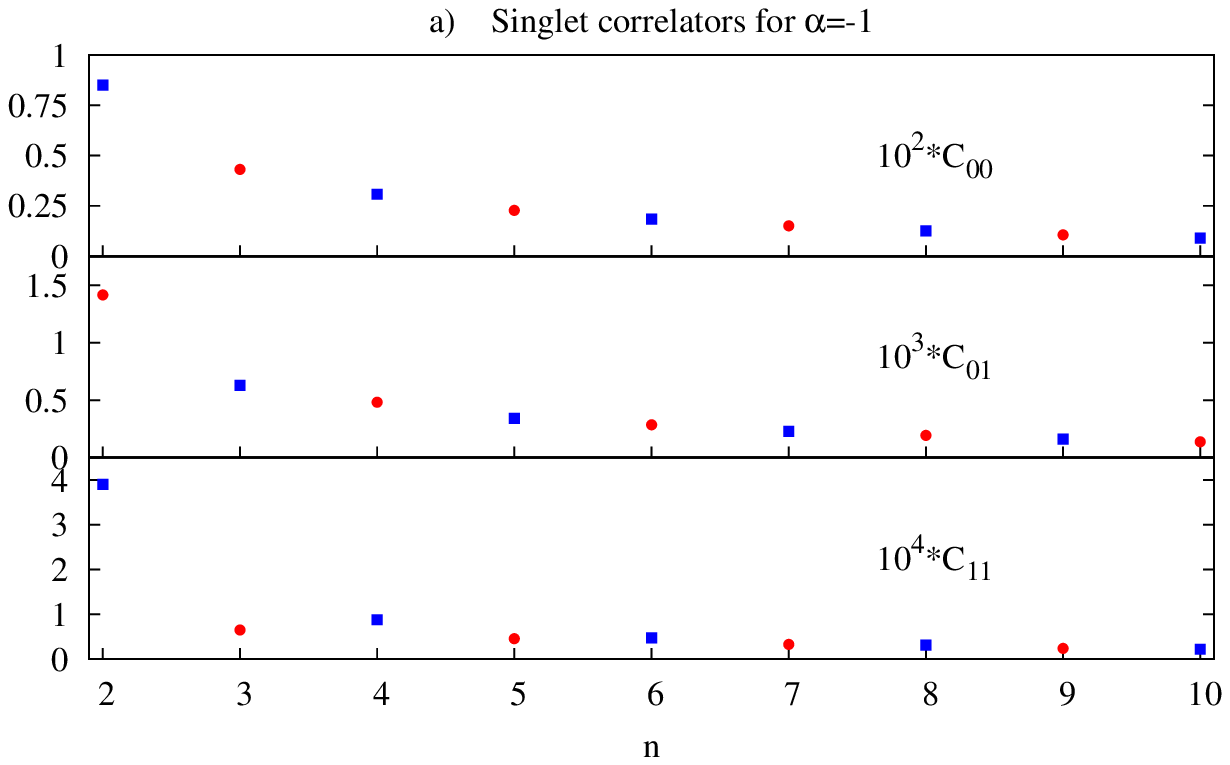}
\includegraphics[width=0.7\linewidth]{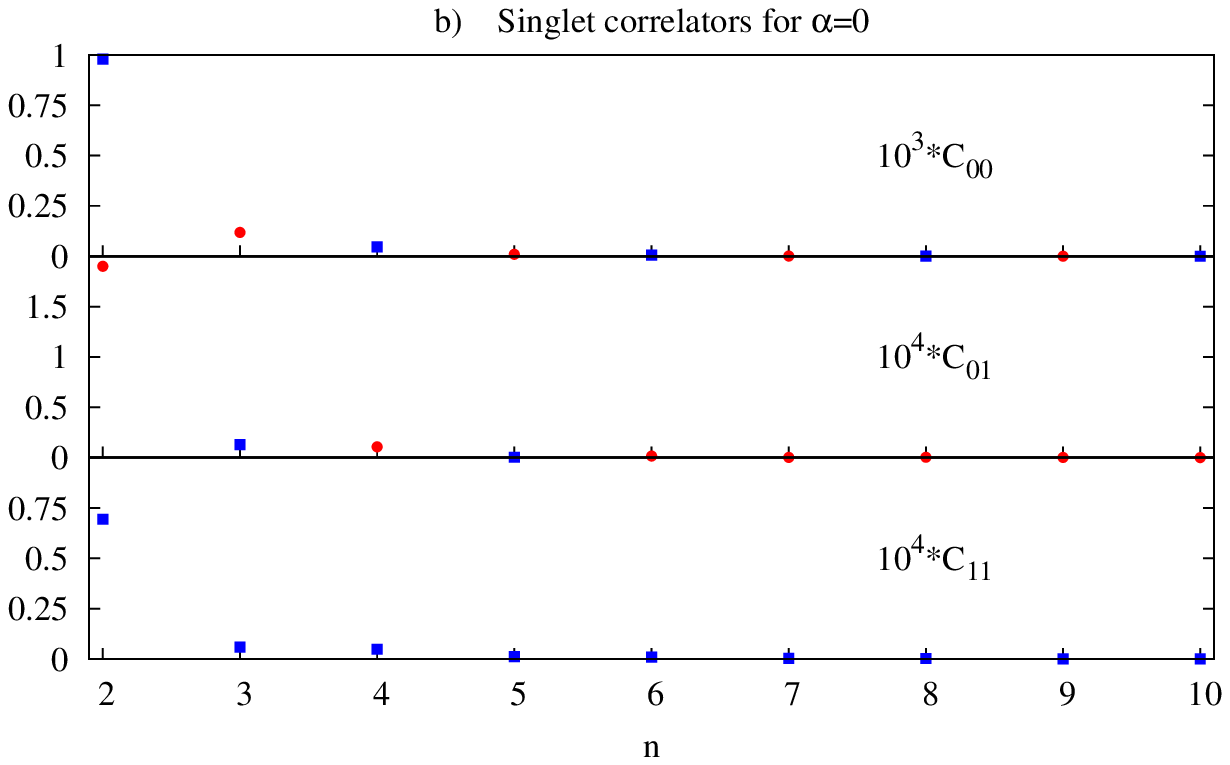}
\end{center}
\caption{Depiction of the connected correlation functions
  $C_{00}(n):=\langle\langle P_0(0)P_0(n)\rangle\rangle$,
  $C_{01}(n):=\langle\langle P_0(0)P_1(n)\rangle\rangle$, and
  $C_{11}(n):=\langle\langle P_1(0)P_1(n)\rangle\rangle$ where $P_0$ and $P_1$
  are projectors onto nearest-neighbour singlets and triplets.  The height and
  colour of the symbols encode the absolute value and the sign: blue squares
  and red circles for plus and minus.  a) The singlet correlators for
  $\alpha=-1$ are relatively small, algebraically decaying and oscillate. b)
  The singlet correlators for $\alpha=0$ are even smaller and exponentially
  decaying. For the shown distances $n\le 10$, $C_{00}$ oscillates. For
  small (large) values of $n$, $C_{01}$ oscillates (is uniform). $C_{11}$ is
  uniform for all $n$.}
\label{graphdimer}
\end{figure}

The dependence of the ground-state energy $e_0$ and of the expectation values
of the projection operators $P_0$, $P_1$ on $\alpha$ is quite smooth -- not to
say boring -- see Fig.~\ref{graphdimeral} a). The true behaviour of the
entanglement entropy would show a singularity at $\alpha=-1$, but only rises
to the finite value of $S=5.2067$ in our treatment due to the finite number of
multiplets. Likewise, the finite string order ${\cal{O}}_2$, see
Fig.~\ref{graphdimeral} b), decreases upon approaching $\alpha=-1$, but does
not drop to exactly 0. We calculated the correlation lengths of singlet
correlation functions from the next-leading eigenvalues of the transfer matrix
$T$. The next-leading positive and negative eigenvalues result in the
correlation lengths $\xi_+$ and $\xi_-$. For approaching $\alpha=-1$ both
lengths increase, in an exact treatment they would diverge. At some
$\alpha_c\simeq -0.6$ the lengths show a crossover. Interestingly, for this
point the papers \cite{OitmaaPB86,OstlundR95} reported a softening of an
excitation gap at momentum $\pi$.
We believe, that this is not a coincidence. In any case, the implication of
the crossover is that for $\alpha<\alpha_c$ singlet correlators will
generically show oscillating asymptotics, see Fig.~\ref{graphdimer} a), and for
$\alpha>\alpha_c$ the asymptotics will be uniformly decaying. However, matrix
elements may be of different orders of magnitude and may obscure this
behaviour at finite distances, see Fig.~\ref{graphdimer} b) where only at
sufficiently large distances the oscillations set in.

The dependence of $\xi_\pm$ on $\alpha$ is not very smooth. This need not be a
numerical artefact. Especially for $\xi_-$ at $\alpha\sim 0.2$ a crossover of
another high-lying negative eigenvalue looks responsible for the
non-monotonous dependence on $\alpha$. Finally, we like to mention that for
$\alpha=1/3$ all singlet correlators are zero.

\section{Conclusion}\label{Conclusion}

In this paper we presented a most economical formulation of general $su(2)$
invariant MPS with arbitrary spin-$s$ in the quantum space. The basic objects
for the calculation of expectation values of nearest-neighbour interactions
and of correlation functions appear as contractions over products of
fundamental tensors of rank 3. The contractions involve internal sums over
`magnetic quantum numbers' which were evaluated explicitly by use of Wigner
calculus. The results of these algebraic calculations were presented in
$su(2)$ invariant manner in terms of $3j$ and $6j$ symbols.

The usefulness of the $su(2)$ invariant formulation was demonstrated at the
example of the spin-1 bilinear-biquadratic quantum chain directly in the
thermodynamical limit. For a large part of the Haldane phase we determined by
variational calculations the ground-state energy, the expectation values of
projection operators on nearest-neighbour singlet and triplet spaces, the
entanglement entropy, string order and singlet operator correlations as well
as their leading correlation lengths.

The achieved accuracy is probably more than acceptable for practical
purposes. Our numerical calculations were done in Maple 13 and resulted in an
accuracy of $2\cdot 10^{-7}$ for the ground-state energy of the strictly
bilinear spin-1 Heisenberg chain. These results were obtained for 5, 4, 2, 1
copies of spin-$\s12$, $\s32$, $\s52$, and $\s72$ multiplets in the auxiliary
space which in total is 46-dimensional.

In future work we want to develop a `normal' program code to do faster
calculations and to deal with larger dimensional auxiliary spaces. Ultimately,
and by making use of the gained experience with the $su(2)$ calculus of MPS,
we intend to tackle tensor-network calculations.

\subsection*{Acknowledgments.}  The authors would like to thank Y. \"Oz and
J. Sirker for valuable discussions. The authors gratefully acknowledge
support by {\em Deutsche Forschungsgemeinschaft} under project {\em
  Renormierungsgruppe KL 645/6-3}. AK acknowledges support by
CNRS and the hospitality of Universit\'e Pierre et Marie Curie where the final
stages of the presented work were carried out.

\end{document}